\documentclass[12pt,eqsecnum,preprint,flushrt]{aastex}
\usepackage{amsmath}
\usepackage{graphicx}
\def\pd#1#2{\frac{\partial #1}{\partial #2}}

\def\prn#1{{\left(#1\right)}}
\def\brk#1{{\left[#1\right]}}

\newcommand{\Del}{\boldsymbol{\nabla}}
\newcommand{\be}{\begin{equation}}
\newcommand{\ee}{\end{equation}}
\newcommand{\qmax}{{ q_\ast }}  
\newcommand{\pmax}{{ p_\ast }}  
\newcommand{\tcool}{{ t_{\rm cool} }} 
\newcommand{\ncool}{{ \alpha_{\rm cool} }} 
\newcommand{\ndiff}{{ \alpha_{\rm diff} }} 

\title{STABILITY OF MAGNETIZED DISKS AND IMPLICATIONS FOR PLANET FORMATION} 

\author{Susana Lizano$^1$, Daniele Galli$^2$, Mike J. Cai$^3$, 
and Fred C. Adams$^{4,5}$} 

\affil{$^1$Centro de Radioastronom{\'i}a y Astrof{\'i}sica, UNAM, Apartado
Postal 3-72, 58089 
Morelia, Michoac\'an, M\'exico} 

\affil{$^2$ INAF-Osservatorio Astrofisico di Arcetri, 
Largo Enrico Fermi 5, I-50125 Firenze, Italy} 

\affil{$^3$Academia Sinica, Institute of Astronomy and Astrophysics, Taiwan}

\affil{$^4$Michigan Center for Theoretical Physics \\
Physics Department, University of Michigan, Ann Arbor, MI 48109} 

\affil{$^5$Astronomy Department, University of Michigan, Ann Arbor, MI 48109} 

\begin{abstract} 

This paper considers gravitational perturbations in geometrically thin
disks with rotation curves dominated by a central object, but with
substantial contributions from magnetic pressure and tension. The
treatment is general, but the application is to the circumstellar
disks that arise during the gravitational collapse phase of star
formation. We find the dispersion relation for spiral density waves in
these generalized disks and derive the stability criterion for
axisymmetric $(m=0)$ disturbances (the analog of the Toomre parameter
$Q_T$) for any radial distribution of the mass-to-flux ratio
$\lambda$.  The magnetic effects work in two opposing directions:
on one hand, magnetic tension and pressure stabilize the disk against
gravitational collapse and fragmentation; on the other hand, they
also lower  the rotation rate making the disk more unstable. For
disks around young stars the first effect generally dominates, so
that magnetic fields allow disks to be stable for higher surface
densities and larger total masses.  These results indicate that
magnetic fields act to suppress the formation of giant planets
through gravitational instability. Finally, even if gravitational
instability can form a secondary body, it must lose an enormous
amount of magnetic flux in order to become a planet; this latter
requirement represents an additional constraint for planet formation
via gravitational instability and places a lower limit on the
electrical resistivity.

\end{abstract} 

\keywords{Magnetohydrodynamics(MHD) -- Stars: formation-- Protoplanetary disks 
-- Planets and   satellites: formation}

\begin{document}
\maketitle

\section{Introduction} 
\label{sec:intro} 

During the gravitational collapse that forms star/disk systems,
magnetic fields are dragged in from the interstellar medium (e.g.,
Galli et al. 2006; Shu et al. 2006).
 Additional fields can be generated by the central star.
Consideration of mean field magnetohydrodynamics (MHD) in these
disks shows that magnetic effects produce substantial departures
from keplerian rotation curves through both magnetic pressure and
magnetic tension (Shu et al. 2007, hereafter S07).  On the other
hand, conservation of angular momentum implies that most of the
material that eventually accretes onto the forming star initially
lands on the disk (Cassen \& Moosman 1981).  As a result, disk
surface densities can be high enough to support gravitational
instability. In the limit of axisymmetric perturbations, the criterion
for gravitational instability is determined by the value of the
parameter $Q_T$,
\be
Q_T \equiv \frac{a \kappa}{\pi G \Sigma} \, , 
\label{qtoomre} 
\ee
where 
$\kappa=\varpi^{-1} [\partial(\varpi^2\Omega)^2/\partial\varpi]^{1/2}$ 
is the epicyclic frequency, $\Omega$ is the angular rotation rate, $a$
is the sound speed, and $\Sigma$ is the surface density (Toomre 1964).
In the presence of magnetic fields, however, the conditions required
for gravitational instability are modified. The principal goal of 
this paper is to generalize the criterion of equation (\ref{qtoomre})
to include the effects of magnetic fields. More specifically, we
derive a generalized stability parameter $Q_M$ that characterizes
magnetized disks.

We note that gravitational instability can play two important roles in
circumstellar disks during the star formation process. If the
instabilities grow into the nonlinear regime, they can produce
secondary bodies within the disk, such as brown dwarfs and giant
planets. If the growing perturbations saturate, the gravitational
torques can lead to redistribution of angular momentum and disk
accretion. Both processes require the onset of gravitational
instability, which is determined by the parameter $Q_M$ derived in
this paper.

The properties and evolution of magnetized disks also depend on the
dimensionless mass-to-flux ratio $\lambda$, defined by
\be 
\lambda \equiv \frac{2\pi G^{1/2} \Sigma_0}{B_{z0}} \, .
\label{lambdef} 
\ee
For example, gravitational collapse requires $\lambda > 1$. As found
by S07 and discussed herein, for realistic magnetic field strengths,
this constraint inhibits the formation of giant planets by
gravitational instability in circumstellar disks. In addition, the 
generalized stability parameter $Q_M$ derived in this paper must be a
function of $\lambda$.

As we show in this paper, the inclusion of magnetic fields leads
to competing effects, some of which inhibit and some of which enhance
gravitational instability and planet formation. However, as outlined
above, magnetic fields will be present within circumstellar disks.
As a result, in order to understand disk physics, one must include
the effects of magnetic fields, and the goal of this paper is to
provide an assessment of these effects.

This paper is organized as follows. We specify the equations of motion
for magnetized disks in Section \ref{sec:basic} and find their
linearized counterparts in Section \ref{sec:linear}. This procedure
leads to the dispersion relation for spiral density waves and the
generalized stability parameter $Q_M$. In Section \ref{sec:numerical}
we present numerical examples and apply the results to the observed
protostellar source Ceph A HW2. The condition $Q_M > 1$ is necessary
for stability and implies a corresponding maximum disk mass, as shown
in Section \ref{sec:maxmass}.  The onset of instability and the
derivation of $Q_M$ can be determined by setting the resistivity
$\eta$ = 0; however, realistic disks have $\eta \ne 0$ and Section
\ref{sec:diffusion} outlines the corresponding effects of magnetic
diffusion.  We then consider giant planet formation in Section
\ref{sec:pformation}. In addition to deriving modified constraints on
planet formation via gravitational instability due to magnetic
effects, we find that magnetic disks require an additional constraint:
The need to remove magnetic flux places a lower bound on the
electrical resistivity $\eta$. Finally, we conclude in Section
\ref{sec:conclude} with a summary and discussion of our results.

\section{Basic Equations}
\label{sec:basic} 

This section specifies the equations of motion for this problem.  We
include the effects of a poloidal magnetic field dragged into the disk
during the gravitational collapse of the the natal cloud that produces
a newly born star/disk system.  This field threads vertically through
the circumstellar disk and is pinched radially inward by viscous
disk accretion.  The accretion in these disks is believed to occur
via the magneto-rotational instability (MRI;  see, e.g., the review
of Balbus \& Hawley 1998).  In fact, an empirical  formulation of
the MRI viscosity in thin disks has been obtained by S07 using
mixing length arguments.

Consider  the evolution of gas and magnetic
field  in a thin axisymmetric, viscously accreting disk of
half-thickness $z_0$, surrounding a young star with mass $M_\star$ at the
origin of a cylindrical coordinate system $(\varpi, z)$.  We denote
the surface density of the disk by $\Sigma$, the radial velocity of
accretion in the plane by $u$, the azimuthal velocity about the $z$
axis by $v$, the component of the magnetic field threading vertically
through the disk by $B_z$, and the radial component of the magnetic
field just above the disk that responds to the radial accretion flow
by $B_\varpi^+$. 
The component of the Lorentz force per unit area in the plane of the disk
can be written as
\be
{\bf f}_{\parallel}=\frac{{\bf B}^+_\parallel B_z}{2 \pi}
-\nabla_\parallel\int \frac{B_z^2}{8 \pi} \, dz ,
\label{lorentz}
\ee
where ${\bf B}^+_\parallel = B^+_\varpi \hat {\bf e}_\varpi
+B^+_\varphi\hat{\bf e}_\varphi$ and $\nabla_\parallel = \hat{\bf e}_\varpi
\partial/\partial\varpi+\hat{\bf e}_\varphi (1/\varpi)\partial/\partial
\varphi$. The two terms in eq.~(\ref{lorentz}) are associated to the 
effects of the magnetic tension and magnetic pressure (see Shu \& Li~1997,
who, however, adopt a different definition of magnetic pressure). In this 
paper we adopt the approximation
\be
\int \frac{B_z^2}{8\pi}\, dz \approx \frac{z_0 B_z^2}{4 \pi},
\ee
where $B_z$ on the r.h.s. is evaluated in the midplane of the disk.
In what follows, we neglect the toroidal component of the magnetic 
field threading the disk. In contrast, the stability of a disk with a purely azimuthal
magnetic field was studied by Lynden-Bell~(1996).

With these specifications, the governing MHD equations 
in cylindrical coordinates include the equation of continuity, 
\be
\pd{\Sigma}{t} + \frac{1}{\varpi} \pd{}{\varpi} (\varpi \Sigma u) 
+ \frac{1}{\varpi^2} \pd{}{\varphi} (\Sigma \varpi v) = 0;
\ee
and the radial and azimuthal components of the equation of momentum,
\be
\Sigma \prn{\pd{u}{t} + u \pd{u}{\varpi} + 
\frac{v}{\varpi} \pd{u}{\varphi} - 
\frac{v^2}{\varpi}} = -\pd{\Pi}{\varpi} - 
\Sigma \prn{\pd{V}{\varpi}}_{z=0} + \frac{B_z B_\varpi^+}{2\pi} - 
\frac{z_0}{4\pi} \pd{B_z^2}{\varpi},
\ee
\be
\Sigma \prn{\pd{v}{t} + u \pd{v}{\varpi} + 
\frac{v}{\varpi} \pd{v}{\varphi}} +
\frac{u v}{\varpi}
= - \frac{1}{\varpi}\pd{\Pi}{\varphi} - 
\frac{1}{\varpi}\Sigma \prn{\pd{V}{\varphi}}_{z=0} - 
\frac{z_0}{4\pi\varpi} \pd{}{\varphi} (B_{\varpi}^{+2} + B_z^2),
\label{azimuth}
\ee
where $V$ is the gravitational potential and $\Pi$ is the gas
pressure integrated over the disk thickness. The vertical component 
of the induction equation takes the form 
\be
\pd{B_z}{t} + \frac{1}{\varpi} \brk{\pd{}{\varpi} 
(\varpi B_z u) + \pd{}{\varphi} (B_z v)} = 
-\frac{1}{\varpi}\pd{}{\varpi}\left[\eta\varpi\left(
\frac{B_\varpi^+}{z_0}-\pd{B_z}{\varpi}\right)\right]
+\frac{1}{\varpi}\pd{}{\varphi}
\left(\frac{\eta}{\varpi}\pd{B_z}{\varphi}\right) \, ,
\ee
where $\eta$ is the electric resistivity.  Note that we ignore the
viscous torque in the azimuthal component of the momentum equation
(\ref{azimuth}) because the viscous timescale is much longer than
the gravitational instability timescale. Also, the azimuthal
component of the magnetic field in the disk that arises from the stretching
of the poloidal field by differential rotation averages to zero in the
vertical integration.

The vacuum fields above and below the disk are treated  using the Green's
function technique (see, e.g., Shu \& Li 1997).  Since ${\bf B}$ is
current-free outside the disk, the magnetic field can be derived from
a scalar potential
\be
{\bf B} = \Del \Psi \, .
\label{defB}
\ee
The condition $\nabla \cdot {\bf B}=0$ then implies that 
$\Psi$ satisfies Laplace's equation 
\be
\nabla^2 \Psi = 0 \mbox{~~for $z \neq 0$} \, ,
\ee
subject to the boundary condition
\be
\left(\pd{\Psi}{z}\right)_{z=0} = B_z  \, .
\ee
The gravitational potential  is given by the stellar and disk contirbutions, $V = -GM_\star/\varpi + V_d$, 
where $V_d$ satisfies Poisson's 
equation for a thin disk,
\be
\nabla^2 V_d = 4 \pi G \Sigma\,\delta(z) \,.
\ee
For simplicity, in the following we assume an isothermal equation of
state, $\Pi=a^2\Sigma$, where $a$ is the local sound speed.

\section{Linearized Perturbation Equations}
\label{sec:linear} 

This section presents a perturbation analysis of the equations of
motion and derives a dispersion relation for spiral density waves in
magnetized disks.  First expand to first order, and use the 
subscript $0$ to denote zeroth order variables and $1$ to denote first
order perturbations.  It is understood that every variable is
evaluated in the midplane (or just above the midplane). We thus look
for solutions that have the Fourier decomposition
\be
F(\varpi, \varphi, t) = 
F_0 (\varpi) + F_1 (\varpi) e^{i(\omega t - m \varphi)} \, , 
\ee
where $\omega$ is a complex frequency and $m$ is a positive integer.
To zeroth order, we assume the disk to be in a state of axisymmetric
radial equilibrium:
\be
\Omega^2\varpi-\frac{a^2}{\Sigma_0}\pd{\Sigma_0}{\varpi}
- \left(\pd{V_0}{\varpi}\right)_{z=0} 
+ \frac{B_{z0}B_{\varpi 0}^+}{2\pi \Sigma_0} 
- \frac{z_0}{4\pi\Sigma_0}\pd{B_{z0}^2}{\varpi} = 0 \, ,
\ee
where $\Omega=v_0/\varpi$. 

The first order equations become
\begin{align}
  &i(\omega - m \Omega)\frac{\Sigma_1}{\Sigma_0} + \frac{1}{\Sigma_0 \varpi} 
\pd{}{\varpi} (\varpi \Sigma_0 u_1) - i m \frac{v_1}{\varpi} = 0,\\
\begin{split}
  &i(\omega - m\Omega) u_1 - 2 \Omega v_1 = 
\frac{\Sigma_1}{\Sigma_0} \prn{- \frac{B_{z0} B_{\varpi0}^+}{2\pi \Sigma_0} + 
\frac{a^2}{\Sigma_0}\pd{\Sigma_0}{\varpi} + 
\frac{z_0}{4\pi\Sigma_0} \pd{B_{z0}^2}{\varpi}} \\
  &-\frac{a^2}{\Sigma_0} \pd{\Sigma_1}{\varpi} - 
\pd{V_1}{\varpi} + \frac{B_{z0} 
B_{\varpi1}^+}{2\pi \Sigma_0} + 
\frac{B_{z1} B_{\varpi0}^+}{2\pi\Sigma_0} - 
\frac{z_0 B_{z1}}{2\pi\Sigma_0} \pd{B_{z0}}{\varpi} 
- \frac{z_0 B_{z0}}{2\pi\Sigma_0} \pd{B_{z1}}{\varpi},
\end{split}\label{radial_pert}\\
  &i(\omega - m \Omega) v_1 + u_1 \frac{\kappa^2}{2\Omega} = 
im \frac{a^2}{\varpi\Sigma_0}\Sigma_1 + im \frac{V_1}{\varpi} 
+ im \frac{z_0}{\pi\varpi \Sigma_0} 
(B_{\varpi0}^+ B_{\varpi1}^+ + B_{z0} B_{z1}),\\
  &i(\omega - m\Omega) B_{z1} + \frac{1}{\varpi} 
\brk{\pd{}{\varpi} (\varpi B_{z0} u_1) -i m B_{z0} v_1} = 
\frac{i}{\varpi}\pd{}{\varpi}\left(\eta\frac{\varpi B_{z1}}{z_0}\right)
-\eta m^2\frac{B_{z1}}{\varpi^2}.
\end{align}
The last two terms in the parenthesis in equation \eqref{radial_pert}
correspond to the thermal and magnetic pressure of the zeroth order
solution, and both of these terms are much smaller than the magnetic
tension term. As a result, we can drop them in the following analysis.
To proceed further, we invoke the WKB approximation and write the
perturbed quantities as
\be
F_1 (\varpi) = {\hat F}_1(\varpi) e^{i k \varpi} \, , 
\ee
where $k$ is the radial wavenumber. We make the further assumption
that $|k|\varpi \gg 1$, i.e., the spiral perturbations are tightly
wrapped.  As a result, we can ignore all derivatives of the amplitude,
or division by $\varpi$, compared to derivatives of the phase.  In
other words, we may replace the radial derivative of a variable with 
multiplication by $ik$ times its amplitude (for simplicity, we will
omit the circumflex accent on the slowly-varying amplitudes).  As an
additional simplification, we assume $1/|k|\varpi$ is of the same
order as $a/\varpi \Omega$.  This specification implies that the
parameter $|k|z_0$ remains order unity. Next we note that the first
order perturbation in surface density and gravitational potential are
related by
\be
\Sigma_1 = - \frac{|k| V_1}{2\pi G} \, ,
\ee
which follows from the asymptotic solution of the Poisson equation for
$V_d$ to the leading WKB order (see Shu~1992).  Following a similar
argument for Poisson's equation for the magnetic potential $\Psi$, 
we also obtain
\be
B_{z1} = - |k| \Psi_1^+ \, ,
\ee
where $\Psi_1^+$ is the value of $\Psi_1$ just above the disk
($\Psi_1$ is an odd function of $z$: if $B_{z1}>0$, then $\Psi_1<0$
for $z>0$ and $\Psi_1>0$ for $z<0$).  From the definition (\ref{defB})
we have
\be
B_{\varpi1}^+ \equiv \pd{\Psi_1^+}{\varpi} = 
-\frac{1}{|k|}\pd{B_{z1}}{\varpi} = -i \frac{k}{|k|}B_{z1} \, .
\ee
Using the WKB approximation in conjunction with the above relationships, 
the first order equations become
\begin{align}
  &i(\omega - m \Omega)\frac{|k|V_1}{2\pi G \Sigma_0} - ik u_1 = 0,\\
\begin{split}
  &i(\omega - m\Omega) u_1 - 2 \Omega v_1 = 
\frac{|k|V_1}{2\pi G \Sigma_0} \left(\frac{B_{z0} B_{\varpi 0}^+}{2\pi\Sigma_0} 
+ ik a^2 \right) - ikV_1 \\
& - i\frac{k}{|k|}(1+|k|z_0)\frac{B_{z0} B_{z 1}}{2\pi \Sigma_0}  
+ \frac{B_{z1} B_{\varpi 0}^+}{2\pi \Sigma_0} - 
\frac{z_0 B_{z1}}{2\pi\Sigma_0} \pd{B_{z0}}{\varpi} ,
\end{split}\\
  &i(\omega - m \Omega) v_1 + u_1\frac{\kappa^2}{2\Omega} = 
- im \frac{a^2|k|V_1}{2\pi G \Sigma_0\varpi} + im \frac{V_1}{\varpi} 
+ im \frac{z_0 B_{z1}}{\pi\Sigma_0\varpi} \left(B_{z0}-i\frac{k}{|k|}B_{\varpi 0}\right),\\
  &i(\omega - m\Omega) B_{z1} + ik B_{z0} u_1 = -\eta |k| \frac{B_{z1}}{z_0}.
\end{align}

We analyze first the case of ideal MHD (we consider the effects
of a non-zero resistivity in Section \ref{sec:diffusion}).  Setting
$\eta=0$ and solving the first and last equations for $u_1$ and
$B_{z1}$, we obtain
\begin{align}
&u_1 = (\omega - m \Omega)\frac{|k|V_1}{2\pi G \Sigma_0k},\\
&B_{z1} = \frac{B_{z0}}{\Sigma_0}\Sigma_1 = - \frac{|k|V_1}{\lambda G^{1/2}}\, , 
\end{align}
where the mass-to-flux ratio is defined through equation (\ref{lambdef}). 
After substituting these results, the remaining equations become 
\begin{align}
  &i(\omega - m\Omega)^2\frac{|k|}{2\pi G \Sigma_0k} - 
2 \Omega \frac{v_1}{V_1} = ik a^2\frac{|k|}{2\pi G \Sigma_0} - 
ik + ik \frac{1}{\lambda^2} (1 + |k|z_0),\\
  &i(\omega - m \Omega) \frac{v_1}{V_1} + 
(\omega - m \Omega)\frac{|k|}{2\pi G \Sigma_0k} 
\frac{\kappa^2}{2\Omega} = \frac{im}{\varpi} 
\brk{1-\frac{a^2|k|}{2\pi G \Sigma_0} - 
\frac{2|k|z_0}{\lambda^2}\left(1-i\frac{kB_{\varpi 0}^+}{|k|B_{z 0}}\right)} \, .
\end{align}
After further algebraic manipulation, the 
leading order dispersion relation takes the form 
\be
(\omega - m \Omega)^2 = \kappa^2 
- 2\pi G \Sigma_0 |k|\epsilon
+ k^2 \Theta a^2 \, ,
\label{dispersion} 
\ee
where we define
\be 
\Theta \equiv 1+\frac{B_{z0}^2z_0}{2\pi \Sigma_0 a^2}\, \mbox{~~~and~~~}
\epsilon \equiv 1 - \frac{1}{\lambda^2} \, . 
\label{Theta_eps}
\ee
In the limit of vanishing magnetic field, this dispersion relation
reduces to the familiar form for spiral density waves in a gaseous
disk (e.g., Shu 1992).  The magnetic field threading the disk
modifies the standard dispersion relation for an unmagnetized disk
by ({\em i}\/) replacing the sound speed $a$ by
the magnetosonic speed $\Theta^{1/2} a = (a^2+v_{A0}^2)^{1/2}$, where
$v_{A0}=B_{z0}^2z_0/2\pi \Sigma_0$ is the
Alfv\'en speed at the disk midplane, and ({\em ii}\/) diluting the effects
of gravity by a factor $\epsilon$ if $\lambda>1$. If $\lambda
< 1$, the right hand side of equation (\ref{dispersion}) is always
positive and no instability occurs.

Although eq.~(\ref{dispersion}) is valid in genaral for a thin disk
with any radial distribution of the mass-to-flux ratio $\lambda$,
it has the same form as the dispersion relation obtained by Shu \&
Li (1997) for the special case of a disk with spatially uniform
$\lambda$ (an ``isopedic'' disk).  In particular, the marginal
stability of isopedic disks with $\lambda=1$ was demonstrated
explicitly by Zweibel \& Lovelace~(1997).  The magnetically modified
Toomre $Q_M$ parameter, which provides the boundary of stability
for axisymmetric ($m=0$) perturbations, is thus given by
\be
Q_M = \frac{\Theta^{1/2}a\kappa}{\pi \epsilon G \Sigma_0} \, .
\label{qmagnetic} 
\ee
Note that the definition of $\Theta$ in equation~(\ref{Theta_eps})
differs slightly from that of Shu \& Li~(1997).  For $Q_M<1$,
perturbations with wavenumber between $k_\pm=k_{\rm max}(1\pm
\sqrt{1-Q_M^2})$ are unstable, with $k_{\rm max}=(\epsilon/\Theta)
k_J$ being the wavenumber of maximum growth, and where $k_J=\pi G
\Sigma_0/a^2$ is the Jeans wavenumber. Since $\epsilon/\Theta <1$, the
effect of the magnetic field is to increase the length scale of the
gravitational instability with respect to the Jeans length scale.

Another important factor that determines $Q_M$ in eq.~(\ref{qmagnetic})
is the epicyclic frequency.  Disks around young stars that have
dragged in magnetic fields from the interstellar medium by gravitational
collapse do not rotate at keplerian speeds because magnetic tension
modifies the force balance equation (see, e.g., eq.~(18) of S07
when magnetic tension dominates over magnetic and gas pressure).
S07 showed that in magnetized disks that are viscously accreting
by the MRI, the rotation curve is subkeplerian by a constant fraction
$f$. In their models, the subkeplerian parameter $f$ is determined
by their equation (73) that states that the magnetic flux brought
in by star formation is conserved and is left behind in the disk.
Thus, for a given mass-to-flux ratio $\lambda $,  the factor $f$
depends on the stellar mass, $M_\star$ (necessary to recover the flux
brought in by star formation), the mass accretion rate, $\dot M_d$,
and the system age, $t_{\rm age}$. For $\lambda \sim 4$, S07 obtained
values of $f$ in the range $0.39$--$0.95$ for disks around low-mass
and massive young stars (see their Table 2).
For subkeplerian disks, the epicyclic frequency is given by 
\be
\kappa = 
f \Omega_K = 
f \left( G M_\ast /\varpi^3 \right)^{1/2} \, . 
\label{subkepler} 
\ee
Therefore, the inclusion of magnetic fields produces competing effects
on the instability parameter $Q_M$: The strong fields enforce
subkeplerian flow, which reduces $Q_M$ and leads to greater
instability. On the other hand, both magnetic pressure and magnetic
tension act to increase $Q_M$ and lead to enhanced stability.

\section{Numerical Values and Observational Application}
\label{sec:numerical} 

To evaluate the numerical values of the quantities derived in the
previous section, we write $\Sigma_0=\mu m_H N_H$, $a=(3kT/2\mu
m_H)^{1/2}$, $\kappa=\Omega={\cal G}$, where $\mu$ is the molecular
weight, $N_H$ is the hydrogen column density, $T$ is the gas
temperature, and ${\cal G}$ is the velocity gradient. The mass-to-flux
ratio thus becomes
\be
\lambda=2.71 \, 
\mu \left(\frac{N_H}{\mbox{$10^{24}$~cm$^2$}}\right)
\left(\frac{B_z}{\mbox{mG}}\right)^{-1},
\ee
\be
\Theta=1+1.15\times 10^{-2} \,
\left(\frac{B_z}{\mbox{mG}}\right)^2
\left(\frac{z_0}{\mbox{AU}}\right)
\left(\frac{N_H}{\mbox{$10^{24}$~cm$^2$}}\right)^{-1}
\left(\frac{T}{\mbox{K}}\right)^{-1},
\ee
\be
\epsilon=1-1.36\times 10^{-1} \,
\frac{1}{\mu^2}
\left(\frac{B_z}{\mbox{mG}}\right)^2
\left(\frac{N_H}{\mbox{$10^{24}$~cm$^2$}}\right)^{-2},
\ee
\be
Q_M=2.12 \, 
\frac{\Theta^{1/2}}{\epsilon \mu^{3/2}}
\left(\frac{{\cal G}}{\mbox{$10^{-2}$~km~s$^{-1}$~AU$^{-1}$}}\right)
\left(\frac{T}{\mbox{K}}\right)^{1/2}
\left(\frac{N_H}{\mbox{$10^{24}$~cm$^2$}}\right)^{-1}.
\ee

It is useful to consider an observed star/disk system where these
results can be applied. The disk around 
the massive protostar Cepheus~A HW2 is threaded by a large scale
magnetic field of strength $B_{z0}\approx 23$~mG at a radius $R_0
\approx 650$~AU, inferred from methanol masers polarization (Vlemmings
et al.~2010). The disk, observed in the continuum and in several
molecular tracers (Patel et al.~2005; Jim\'enez-Serra et al.~2007,
2009) is seen almost edge-on, with the magnetic field roughly
perpendicular to the disk and inclined with respect to the line of
sight by an angle $i\approx 73^\circ$ (Vlemmings et al.~2010). This
inclination is in agreement with the high aspect ratio of the CH$_3$CN
and SO$_2$ integrated emission, corresponding to inclinations of $\sim
68^\circ$ and $\sim 79^\circ$, respectively (Patel et al.~2005;
Jim\'enez-Serra et al.~2007). The disk shows a velocity gradient
${\cal G} \sin i \approx 6$~km~s$^{-1}$ over $0.5^{\prime\prime}$
(Patel et al.~2005), corresponding to a radial range of $\sim 360$~AU
at the distance of 725~pc. Assuming quasi-keplerian rotation, the
velocity gradient at the radius where the magnetic field has been
measured implies that ${\cal G} \sin i \approx 7\times
10^{-3}$~km~s$^{-1}$~AU$^{-1}$. The inferred mass of the central star
then becomes $M_\star\approx 15 \sin^{-2} i$~$M_\odot$.  The gas
temperature in the outer regions of the molecular disk is $T\approx
250$~K (Jim\'enez-Serra et al. 2009), corresponding to an isothermal
sound speed of $a \approx 1.8\, \mu^{-1/2}$~km~s$^{-1}$.  Assuming
that the disk is thermally supported, the scale height $z_0 \approx
a/\Omega$ at $R_0$ is then $z_0 \approx 260\,\mu^{-1/2} \sin
i$~AU. If the density in the methanol maser region has the typical
value $n_H \approx 10^9$~cm$^{-3}$, the hydrogen column density is
$N_H \approx 7.8\times 10^{24}\,\mu^{-1/2} \sin i$~cm$^{-2}$.

Assuming $i\approx 73^\circ$ for consistency with the magnetic field
strength determination, and $\mu=2.33$, we obtain $\lambda\approx 1.3$
(in agreement with the estimate of Vlemmings et al.~2010),
$\Theta\approx 1.8$, $\epsilon \approx 0.45$.  With these values, the
combined effects of magnetic pressure and tension increase the value
of the Toomre $Q_T$ parameter by a factor of $\Theta^{1/2} / \epsilon
\approx 3$, reinforcing the stability of the disk from a marginally
stable $Q_T\approx 1.4$, to a safer $Q_M\approx 4.3$. 

\begin{figure} 
\figurenum{1}
{\centerline{\epsscale{0.90} \plotone{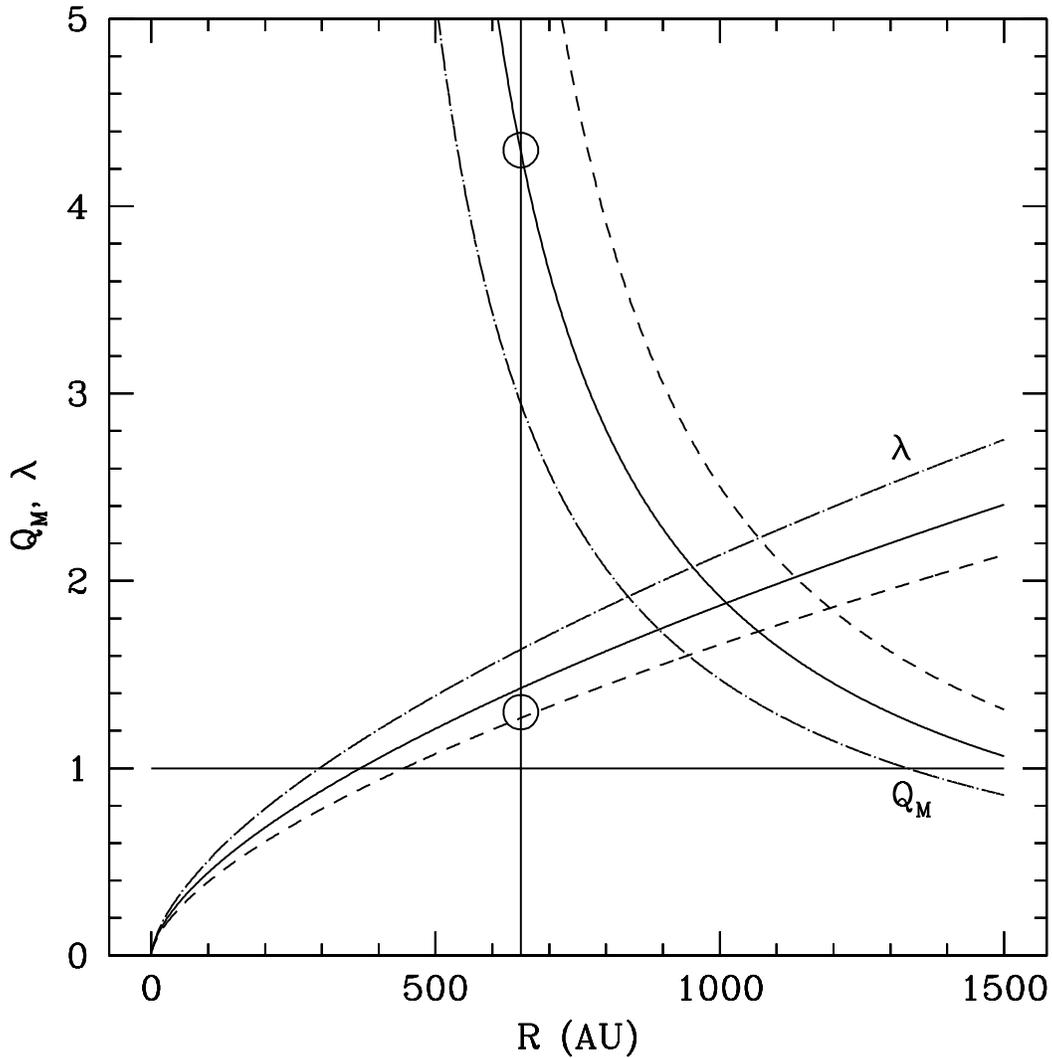}}} 
\figcaption{Values of the dimensionless mass-to-flux ratio $\lambda$ and 
magnetically modified Toomre $Q_M$ as function of disk radius $r$ for 
the case of the circumstellar disk around the massive protostar Cepheus~A HW2
($M_\star=16$~$M_\odot$, $\dot M=10^{-4}$~$M_\odot$~yr$^{-1}$). The
curves correspond to different values of $f$, where $f$ = 0.91
(dashes), $f$ = 0.92 (solid), and $f$ = 0.93 (dot-dashes).  The small
circles indicate the values of $\lambda$ and $Q_M$ derived in the text
at $R=650$~AU.}

\label{fig:CephA}
\end{figure}

Figure \ref{fig:CephA} shows the theoretical dimensionless 
mass-to-flux ratio $\lambda$ (eq. \ref{lambdef}) 
and the magnetic parameter $Q_M$ (eq. \ref{qmagnetic})
as a function of radius $R$. The disk is unstable to fragmentation
at large radii where $\lambda > 1$ and $Q_M < 1$. The sound speed 
$a$, $\lambda$, and the column density 
$\Sigma$ (required to compute $Q_M$) were calculated
using the profiles of magnetized disks in S07, for a system with
a central star, $M_\star=16$~$M_\odot$, and mass accretion rate, 
$\dot M=10^{-4}$~$M_\odot$~yr$^{-1}$, which
correspond to the expected values in the circumstellar disk around the 
massive protostar protostar Cepheus~A HW2. The curves correspond to 
different values of the subkeplerian constant $f$.  The small
circles indicate the values of $\lambda$ and $Q_M$ derived in the text
at $R=650$~AU and the best fit corresponds to $f = 0.92$.  

The numerical values discussed here are certainly subject to
considerable uncertainties. We estimate that the  calculated $Q_M$
could vary in the range from a minimum of 1.3 to a maximum of 7.3,
depending on the uncertainty on the disk's inclination, column
density, magnetic field and velocity gradient.  While it is conceivable
that this rather large range will be reduced by future observations,
this example serves as a ``proof of concept'' to illustrate the
stabilizing effect of magnetic fields in circumstellar disks, and
their importance for the process of planet formation, addressed in
the following sections.

\section{Maximum Disk Mass}
\label{sec:maxmass} 

In this section, we estimate the maximum disk mass that can remain
stable against gravitational collapse in the presence of magnetic
effects. 
We first manipulate the critical stability equation into the
form
\be
1 + \left(\frac{\pi G \Sigma_{\rm max}} {a \kappa} \right)
\left(\frac{2 \beta }{\lambda^2}\right) = 
\left(\frac{\pi G \Sigma_{\rm max}}{a \kappa} \right)^2  
\left(1-\frac{1}{\lambda^2}\right)^2 Q_M^2 \, ,
\label{crit_stab}
\ee
where we defined the disk scale height parameter 
$\beta = \kappa z_0 /a$. 

If we set $Q_M = 1$ and consider both the flux-to-mass ratio 
$\lambda$ and  $\beta$ to be a known functions of radius, we can 
solve equation~(\ref{crit_stab}) for the critical (maximum)
surface density profile 
\be
\Sigma_{\rm max}=\left(\frac{a\Omega_K}{\pi G}\right)f{\cal M}\, ,
\label{Scrit}
\ee
where
\be
{\cal M}=\frac{\lambda^2}{(\lambda^2-1)^2}
\left\{\beta+\left[\beta^2+(\lambda^2-1)^2\right]^{1/2}\right\} \, .
\label{defM}
\ee
The first factor in
equation~(\ref{Scrit}) represents the critical surface density in the
absence of magnetic effects, the second factor represents the reduction
of Keplerian rotation by magnetic tension, and the third represents the
increase in the critical mass due to magnetic support. Notice that 
${\cal M}\rightarrow 1$ in the limit of unmagnetized disk 
($\lambda \rightarrow \infty$). In general, the dimensionless
quantities $f$, $\lambda$, and $\beta$ will be functions of the radial
coordinate $\varpi$ in the disk.  However, one can choose constant
representative values to get an idea of how disk stability depends 
on these quantities. In particular, for a thermally supported
keplerian disk, the scale height $z_0 \approx a/\kappa$, and thus
$\beta\approx 1$.  

\begin{figure} 
\figurenum{2}
{\centerline{\epsscale{0.90} \plotone{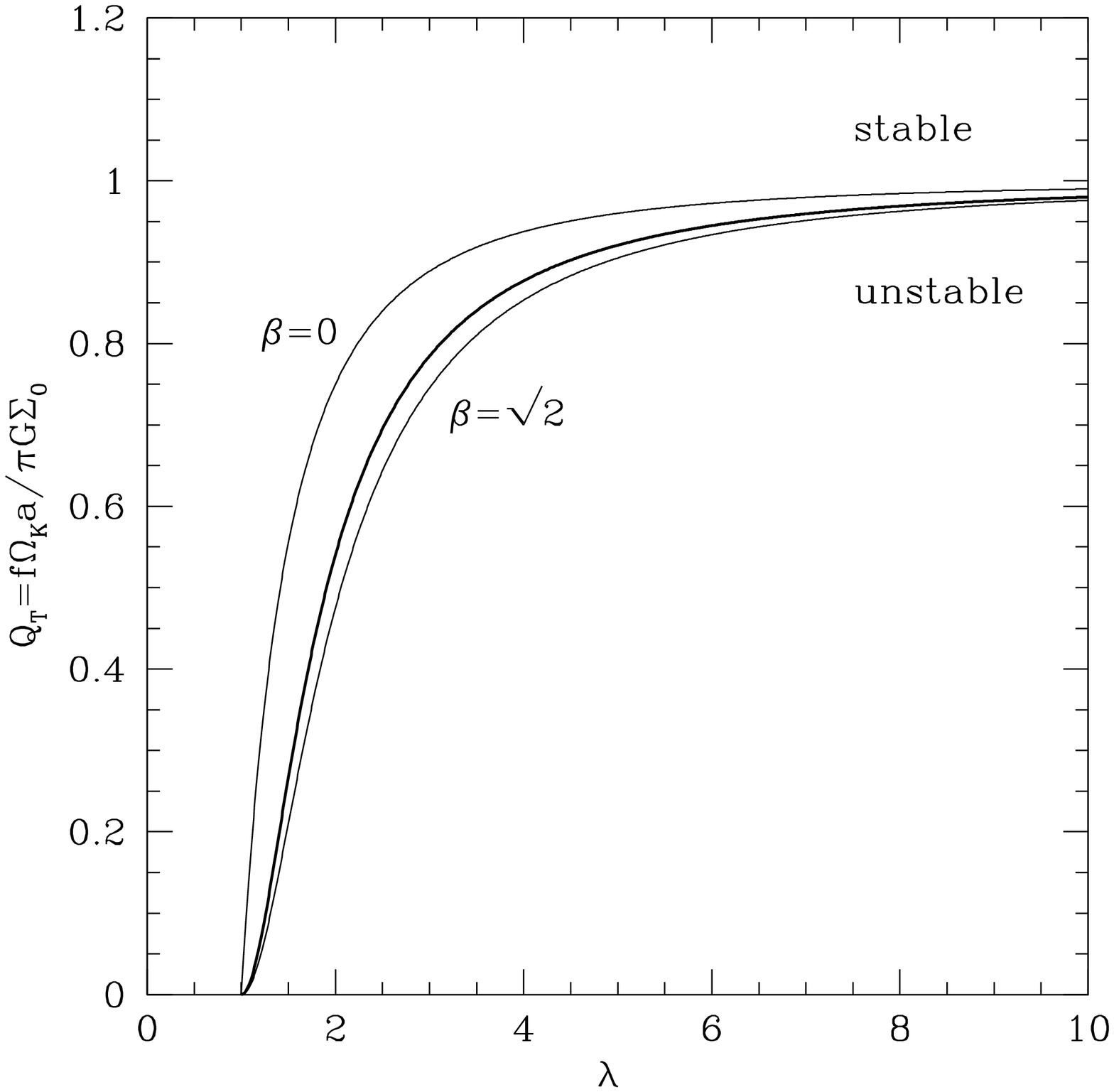} } } 
\figcaption{Curves of $Q_M=1$ in the $\lambda$--$Q_T$ parameter space
for different values of the scale height parameter: $\beta=1$ (thick
curve), $\beta=0$ and $\beta=\sqrt{2}$ (thin curves).  The disk is unstable
in the region below the curves and stable above.  For $f=1$, the line $Q_T = 1$
defines the stability boundary in the absence of magnetic effects. 
Note that magnetic fields act to increase the region for stability. }
\label{fig:qlambda}
\end{figure}

Figure \ref{fig:qlambda} illustrates the manner in which the magnetic 
field affects the stability of a circumstellar disk against 
gravitational perturbations, as expressed by equation (\ref{Scrit}).
Each curve in the $\lambda$--$ Q_T$ plane, where $Q_T = 
a f \Omega_K/\pi G \Sigma_0$ represents the condition $Q_M=1$ for different 
values of the disk scale height parameter $\beta$. The thick curve corresponds to
$\beta=1$ and the thin curves correspond to $\beta=0$ and $\beta=\sqrt{2}$, 
which is the maximum value allowed for a  thermally supported disk (S07).  For each value of $\beta$,
the portion of the plane below the corresponding curve represents the
region of parameter space for which disks are unstable in the presence
of magnetic effects ($Q_M<1$). Above the curve, $Q_M$ is larger than
unity, and the disk is stable. The effects of magnetic pressure and
magnetic tension make a disk more stable compared to its unmagnetized
counterpart, whereas the effects of subkeplerian rotation destabilize
the disk.  For $f=1$, the line $Q_T = 1$ defines the boundary for stability in the
absence of magnetic effects.
The same curves also show the value of the inverse of the
function ${\cal M}$ defined in equation~(\ref{defM}).

Next we define the benchmark disk mass $M_{\rm max}$ integrating
the critical surface density when there are not magnetic effects, 
\be
M_{\rm max} \equiv \int_{R_\ast}^{R_d} 2 \varpi d\varpi 
\frac{a \Omega_K}{\pi G} \, . 
\label{maxzero}
\ee
As shown by many authors (starting with Adams et al. 1988), if one
uses the observed spectral energy distributions of T Tauri star/disk
systems to specify the radial distribution of temperature, and hence
the sound speed profile $a(\varpi)$, the mass scales resulting from
equation (\ref{maxzero}) lie in the range $M_{\rm max}$ = 0.3 -- 1
$M_\odot$, i.e., masses comparable to those of the central stars. 
Note that this benchmark mass scale is calculated by assuming that
$Q_T$ = 1 throughout the disk, and thus represents an upper limit on
the disk mass that can be stable. In practice, the Toomre parameter
depends on radius, $Q_T = Q_T (\varpi)$, so that much of the disk will
have larger $Q_T$ and hence smaller surface density than used in this
exercise. Considerations of global stability show that the maximum
disk mass is lower, e.g., the maximum disk mass that is stable to a
class of $m$ = 1 modes is given by $M_d / (M_\ast + M_d) = 3/4\pi$
(Shu et al. 1990). Nonetheless, this mass scale of equation 
(\ref{maxzero}) provides an interesting benchmark. For the case of 
magnetized disks, the maximum mass is increased, as shown below.

With the inclusion of magnetic effects, the maximum disk mass that is
stable to gravitational perturbations takes the form
\be
M_{{\rm max}, M} \equiv \int_{R_\ast}^{R_d} 2 \varpi d\varpi  
\left( \frac{a \Omega_K}{\pi G} \right) \, f \, 
\left(\frac{\lambda}{\lambda^2 - 1}\right)^2 \left\{ \beta + 
\left[ \beta^2 + (\lambda^2 - 1)^2 \right]^{1/2} \right\} \, , 
\label{maxmass} 
\ee
where, in general, $f$, $\lambda$, and $\beta$ are functions of $\varpi$. 
In the simplest case, however, we can take these parameters to be constant, 
with representative values. In this case, the maximum allowed disk mass 
takes the form 
\be
M_{{\rm max}, M} = f {\cal M} M_{\rm max} = M_{\rm max} \, f \,  
\left(\frac{\lambda}{\lambda^2 - 1}\right)^2 \left\{ \beta + 
\left[ \beta^2 + (\lambda^2 - 1)^2 \right]^{1/2} \right\} \, . 
\label{magmaxmass} 
\ee
This result differs from the field-free case by a factor ${\it F}$ =
$f {\cal M}$.  For example, for the choice of parameters $(f,\lambda,
\beta)$ = (0.9, 1.3, 1), roughly corresponding to the observed values
for the protostellar source Cepheus A HW2 considered in the previous
section, the factor is ${\it F} \approx$ 8 .  For somewhat larger mass to
flux ratios $\lambda$ = 2, the factor ${\it F} \approx 1.8$.  As a
result, magnetic fields can produce a significant increase in the
allowed masses of stable circumstellar disks. In the limit $\lambda \to
\infty$, we expect $f \to 1$ so that ${\it F} \to 1$.

It is important to keep in mind that the condition $Q_M > 1$ is
necessary for the disk to remain stable to axisymmetric perturbations
only. An important result from spiral density wave theory is that
disks which are stable to $m$ = 0 perturbations can still be unstable
to spiral perturbations with $m \ne 0$. As a result, equation
(\ref{maxmass}) represents an upper limit to the maximum stable disk
mass in these systems. Similarly, equation (\ref{maxzero}) represents
an upper limit for the maximum stable disk mass in the absence of
magnetic fields. The actual maximum disk mass is smaller than these
benchmark scales by a factor of $\sim 2$ (e.g., Adams et al. 1989, Shu
et al. 1990).

\section{Inclusion of Magnetic Diffusion} 
\label{sec:diffusion} 

If we include the effects of magnetic diffusion, $\eta \ne 0$, 
then the dispersion relation derived above is modified to take 
the form 
$$
(\omega - m \Omega)^2 
\left( \omega - m \Omega - i \eta \frac{|k|}{z_0} \right) =  
(\omega - m \Omega) \left[ \kappa^2 + k^2 a^2 
\prn{1 + \frac{B_{z0}^2z_0}{2\pi \Sigma_0 a^2}} - 
2\pi G \Sigma_0 |k| \left(1-\frac{1}{\lambda^2}\right) \right] 
$$
\be
-i\eta\frac{|k|}{z_0}\left(\kappa^2+k^2 a^2-2\pi G\Sigma_0|k|\right)  
-\eta k^2\frac{B_{z 0}B_{\varpi 0}^+}{2 \pi \Sigma_0 z_0} \, .
\ee
A basic analysis of this expression shows that the dispersion 
relation has no solutions for which the frequency $\omega$ is 
purely real. As a result, $\omega$ must be complex and can be 
written in the form  
\be
\omega = \omega_0 + i \gamma \, , 
\ee
where both $\omega_0$ and $\gamma$ are real. In addition, 
since the parameter $\eta$ is expected to be small, the 
imaginary part of the frequency is expected to be much 
smaller than the real part, $|\gamma| \ll \omega_0$. 
It is useful to define the functions 
\be
A(k) \equiv \kappa^2 + k^2 a^2 - 2\pi G \Sigma_0 |k| \, , 
\ee
and 
$$
B(k) \equiv \kappa^2 + k^2 a^2 
\prn{1 + \frac{B_{z0}^2z_0}{2\pi \Sigma_0 a^2}} - 
2\pi G \Sigma_0 |k| \left(1-\frac{1}{\lambda^2}\right) 
$$
\be
= A(k)+\frac{2\pi G\Sigma_0}{\lambda^2}|k|(1+|k|z_0) \, . 
\ee 
We also specialize to the case of axial symmetry ($m$ = 0)
so that the dispersion relation has the form 
\be
\omega^2 \left(\omega - i \eta \frac{|k|}{z_0} \right) = 
\omega B(k) - i \eta \frac{|k|}{z_0}A(k) - \eta k^2 b^2 \, , 
\ee
where we have also defined 
$b^2 \equiv B_{z0} B_{\varpi 0}^+/(2 \pi \Sigma z_0)$.
Solving for the real and imaginary parts of the dispersion 
relation, and working to leading order, we find 
\be
\omega_0 \left( \omega_0^2 + 2 \gamma 
\eta\frac{|k|}{z_0} \right) = \omega_0 B(k) - \eta b^2 k^2 \, , 
\ee
and 
\be
\omega_0^2 \left( 3 \gamma - \eta \frac{|k|}{z_0} \right) = 
\gamma B(k) - \eta \frac{k}{z_0} A(k) \, . 
\ee
To leading order, $\omega_0^2 = B(k)$, so the imaginary 
part of the frequency (the growth rate) takes the simple 
form 
\be
\gamma=\eta\frac{|k|}{2 z_0}\left[1-\frac{A(k)}{B(k)}\right] \, . 
\label{decay}
\ee
Since $B(k) \ge A(k)$, with equality only in the limit $|k| \to 0$ or
$\lambda \to \infty$, the growth rate is always positive, so that the
solutions decay like $\exp(-\gamma t)$. 

This decay of the perturbation solutions is expected, in general
terms, because the disk must spread for the case $\eta \ne 0$.
However, the specific form $\gamma \sim \eta |k| / z_0$ is less
obvious.  Nonetheless, this result can be derived by solving the
magnetic induction equation in the limit of an infinitesimally thin
disk (and this calculation is carried out in the Appendix). 

\section{Giant Planet Formation by Gravitational Instability} 
\label{sec:pformation} 

Gravitational instabilities in circumstellar disks can, in principle,
lead to the formation of giant planets, or somewhat larger secondary
bodies such as brown dwarfs (e.g., Boss 2001).  In disk systems with
significant magnetic support, however, the formation of secondaries is
highly suppressed, as outlined in this section.  The formation of
secondary bodies requires both the onset of gravitational instability
and sufficiently short cooling time scales (e.g., Gammie 2001). The
required compromise between these two constraints is modified by
sufficiently strong magnetic fields and is calculated in this section
(compare with Rafikov 2005). Even when these two constraints are met,
so that gravitational instability could in principle produce
secondaries, the magnetic flux problem remains; in other words, the
forming protoplanet must reduce it's magnetic flux in order to
contract to planetary sizes. This latter issue is also addressed
below.

We first consider the coupled constraints of gravitational instability
and sufficient cooling.  In order for gravitational perturbations to
grow, the stability parameter $Q_M$ must be sufficiently small.
Although the growth of axisymmetric instabilities requires $Q_M \le
1$, spiral gravitational instabilities can grow in many star/disk
systems under the weaker condition $Q_M < \qmax \approx 2$ (e.g.,
Adams et al. 1989), where the parameter $\qmax$ defines the required
threshold. This constraint, a necessary condition for gravitational
instability, takes the form 
\be
Q_M = \frac{\Theta^{1/2} a \kappa }{\pi \epsilon G \Sigma} 
< \qmax \, , 
\label{gravcon} 
\ee
where the dimensionless parameters $\Theta$ and $\epsilon$ are defined
above (equation [\ref{Theta_eps}]). 

The survival of gravitational instabilities requires that the cooling
time $\tcool$ is sufficiently short, and this condition takes the 
form $\Omega_K \tcool < \ncool$, where the value of $\ncool \sim 3$.
Note that this constraint uses the Keplerian rotation rate $\Omega_K$;
any departures from Keplerian can be incorporated into the value of
$\ncool$.  Following previous authors (e.g., Gammie 2001), we specify
the form of the cooling time according to
\be
\tcool = \frac{\Sigma a^2}{\gamma - 1} \cdot
\frac{\tau + 1/\tau}{2 \sigma T^4} \, , 
\label{tcool} 
\ee
where $\tau$ is the optical depth of the disk, $\gamma$ is the adiabatic
index,  and $T$ is the temperature. 
The cooling time constraint then takes the form 
\be
\frac{\Omega_K \Sigma a^2}{\sigma T^4} \, < \pmax \, , 
\label{coolcon} 
\ee
where we have defined the dimensionless parameter 
$\pmax \equiv 2 \ncool (\gamma-1)$/$(\tau + 1/\tau)$. 
Note that for typical values $\ncool=3$, $\gamma=5/3$, 
and $\tau=1$, the parameter $\pmax$ = 2. 

By combining the constraints of equations (\ref{gravcon}) and
(\ref{coolcon}), one can show that the disk temperature --- at the
radial location where secondary formation could occur -- obeys the
ordering
\be
\left( \frac{\mu}{k} \right)^3 
\left( \frac{\pi G \Sigma}{{\cal F} \kappa} \right)^6 > T^3 > 
\frac{k}{\mu} \cdot \frac{\Omega_K \Sigma}{\pmax \sigma} \, , 
\label{ordering} 
\ee
where we have used the relation $T = \mu a^2/k$,  and 
\be
{\cal F} \equiv \frac{1}{\qmax^2} 
\left( \frac{\lambda}{\lambda^2 - 1} \right)^2 
\left\{ \beta + \left[ \beta^2 + \qmax^2 
(\lambda^2 - 1)^2 \right]^{1/2} \right\} \, . 
\label{calfdef} 
\ee
By eliminating the temperature using the end points of equation
(\ref{ordering}), one finds the following constraint on the surface
density 
\be
\left( \frac{\mu}{k} \right)^4 
\left( \frac{\pi G \Sigma}{{\cal F} \kappa} \right)^6 > 
\frac{\Omega_K \Sigma}{\pmax \sigma} \qquad \Rightarrow \qquad 
\Sigma \, > \, \left( \frac{f {\cal F}}{\pi G} \right)^{6/5} 
\left( \frac{k}{\mu} \right)^{4/5} 
\frac{\Omega_K^{7/5}}{(\pmax \sigma)^{1/5}} \, .  
\label{sigmacon} 
\ee
If we specialize to the case of a solar-mass star, with 
hydrogen gas ($\mu = 2 m_P$), and $\pmax$ = 2, the surface 
density constraint takes the form 
\be
\Sigma > (3 \times 10^5 \, {\rm g} \, \, {\rm cm}^{-2} ) 
\left( f {\cal F} \right)^{6/5} \, 
\left( \frac{\varpi}{1 \, {\rm AU}} \right)^{-21/10} \, .   
\label{sigmaconnum} 
\ee 
For comparison, the Minimum Mass Solar Nebula (MMSN) is expected to have a
much lower surface density at $\varpi$ = 1 AU, where $\Sigma_1 \sim
4500$ g cm$^{-2}$ or smaller (e.g., Kuchner 2004). If the surface
density profile is a negative power-law with index $p < 2.1$, then
gravitational instability can operate in the outer disk. For the
parameter values listed here, planet formation via gravitational
instability can only take place at radii exceeding the bound
\be
{\varpi}
> \left[ 67 \left( f {\cal F} \right)^{6/5} \right]^{10/(21-10p)} \, {\rm AU}
\approx 1100 \left( f {\cal F} \right)^{2} \,  {\rm AU} ,  
\label{radialscale} 
\ee
where the final equality specializes to the index $p$ = 3/2, often
used for the MMSN. Recall that the parameter $f$ determines the degree
to which the disk is subkeplerian and ${\cal F}$ is defined through
equation (\ref{calfdef}). Although the result is a somewhat
complicated function of disk parameters (especially those that
characterize the magnetic field), the required radius is always 
large. Planet formation via gravitational instability can only 
take place in the outer regions of large disks. 
We note that most disks surrounding low-mass stars have 
outer radii of order 100 AU and hence will not generally 
extend out to 1100 AU where giant planets could form. In 
addition, any planets that form at such large distances 
would have trouble migrating inward to the locations where 
(most of) the current sample of extrasolar planets resides 
($a < 10$ AU). This channel of planet formation, via 
gravitational instability, is thus expected to have a 
limited impact on current observations

The MMSN, with its benchmark mass scale $M_d \approx 0.05 M_\odot$,
typically has an outer radius of only 30 AU. For disks that have the
same form for their surface density as the MMSN, but extend out to the
radial scales of equation (\ref{radialscale}), the corresponding mass
is larger, about $M_d \approx 0.2 M_\odot$ $( f {\cal F} )^{1/2}$.
Notice that the constraints on radial location are weaker if the index
$p$ for the surface density profile is smaller. However, the
constraint on the mass scale is nearly the same. 


Next we consider the magnetic flux problem in the context of forming
secondary bodies. Even if the disk is heavy enough to become unstable,
and the cooling time is short enough to allow contraction, magnetic
flux must be transferred out of the region where the planet forms.
The length scale $\Delta R$ that contains enough mass to form a Jovian
planet is given by the integral over a disk annulus that contains the
planet mass, i.e.,
\be
2 \pi \int_{R_1}^{R_2} \Sigma \varpi d \varpi = m_P,
\ee
where $m_P \approx 1 \, m_J = 2 \times 10^{30}$ g.  For a MMSN disk
model, where $\Sigma = \Sigma_1 (\varpi / R_1)^{-3/2}$, where $R_1$ = 1
AU, this length scale is given by 
\be
\Delta R = R_2 - R_1 \approx 
\frac{m_P}{2 \pi \Sigma_1 R_1} \left( \frac{\varpi}{1 \, {\rm AU}} \right)^{1/2}
\approx 0.31 \, {\rm AU} \, 
\left( \frac{\varpi}{1 \, {\rm AU}} \right)^{1/2} \, 
\approx 10 \, {\rm AU} \, ,
\ee
where $\varpi$ is the radial location of the forming planet and where
we have chosen the nominal location $\varpi$ = 1100 AU as a reference
radius (see equation [\ref{radialscale}]). Gravitational instability
can potentially form fragments with Jovian mass and this length scale
(10 AU), which is much smaller than the Hill's radius, $R_H$ = 
$\varpi (m_P/3 M_\ast)^{1/3} \approx 76$ AU at this location. As a
result, the fragment can remain gravitationally bound.

Let's now consider flux freezing. The initial fragment is threaded
with a magnetic field $B_0$, which can be written in terms of the mass
to flux ratio $\lambda$ through the relation $B_0 \approx$ $2 \pi
G^{1/2} \Sigma / \lambda$. If we use the required values of $\Sigma$
from the above analysis, and for $\lambda$ of order unity, the initial
field strength is about $B_0 \approx 10^{-4}$ G at the large radii
where gravitational perturbations can grow.  To form a giant planet,
the fragment must contract to planetary size scales $R_P \sim$
$10^{10}$ cm. Flux freezing implies that 
\be
B_0 R_0^2 \approx B_P R_P^2 \, , 
\ee
where $B_P$ is the magnetic field strength expected on the planetary
surface. For the values $R_0$ = 10 AU and $R_P$ = few $R_J$, the
surface field strength would be $B_P \approx 2 \times 10^4$ G. As a
result, planet formation requires flux freezing to be compromised.

In these disks, the resistivity $\eta \ne 0$ plays the role of a
diffusion constant for the magnetic field. The time required for the
resistivity to remove magnetic field from a region of size $\ell$ is
thus the diffusion time $t_{\rm diff}$ and is given approximately by
\be 
t_{\rm diff} \sim \ell^2 / \eta \, . 
\label{difftime} 
\ee 
In this case, $\ell \sim (z_0 \Delta R)^{1/2} \sim 45$ AU, where
the disk  half thickness is given by $z_0 = A \varpi$, and the aspect ratio
is taken to be  $A= 0.1 (\varpi/100 {\rm AU})^{1/4}$. The value of the resistivity is 
$\eta \approx 2 \times 10^{20}$ cm$^2$ s$^{-1}$, required to dissipate enough
magnetic flux to meet the constraints posed by  measurements of paleomagnetism in
meteorites (Shu et al. 2006).
With these values, the diffusion time is only $t_{\rm diff} \approx$ 
74 yr. To leading order, the magnetic field strength decreases
exponentially with time, with timescale $t_{\rm diff}$.  In order to
decrease the field strength by the required factor of $10^4$, the
system needs $4\log 10 \approx 9.2$ timescales, or about 680 yr.  The
expected values of the resistivity are thus high enough to allow the
magnetic field to diffuse out of the protoplanet on a short timescale
(680 yr), much shorter than the orbital timescale at the large radii
where giant planets could form. Nonetheless, the required flux loss 
places a lower bound on the resistivity, i.e., the diffusion time scale 
$t_{\rm diff}$ from equation (\ref{difftime}) must be shorter than the 
cooling time.  This constraint can be written in the form 
\be
\eta > \ndiff \Omega \ell^2 \approx \ndiff \Omega 
\left( \frac{m_P A}{2 \pi \Sigma} \right) \, \sim 2.5 \times 10^{18} {\rm cm^2 s^{-1} }, 
\label{etacon} 
\ee
where $\ndiff$ is a dimensionless parameter of order unity and the numerical estimate
was obtained for the density of the MMSN at 1100 AU. 

To conclude this section: The loss of magnetic flux is crucial to the
planet formation process, but this flux can be removed with reasonable
values of the resistivity $\eta$, which must obey the constraint of
equation (\ref{etacon}).  The bottleneck in the planet formation
process is given by the coupled constraints of gravitational
instability and short cooling timescales, where these constraints are
modified significantly by the presence of magnetic fields. As a
result, we expect giant planet formation via gravitational instability
to be highly suppressed in these systems.

\section{Conclusion} 
\label{sec:conclude} 

This paper has generalized the dispersion relationship for spiral
density waves in circumstellar disks to include the leading order
effects of magnetic fields for any radial distribution of the mass-to-flux ratio
$\lambda$ (equation [\ref{dispersion}]). This
procedure results in a generalized version of the stability parameter
(denoted here as $Q_M$) for gravitational instabilities in magnetized
disks, where $Q_M$ is given by equation (\ref{qmagnetic}).  Magnetic
fields produce competing effects regarding the stability of disks to
gravitational perturbations: The increased pressure support and
magnetic tension lead to greater stability; however, these same forces
lead to subkeplerian rotation curves, which in turn lead to greater
instability. The supporting terms generally dominate (see Figure
\ref{fig:qlambda}), so that magnetic effects lead to an overall
suppression of gravitational instabilities. In particular, there
exists a maximum disk mass that is stable to gravitational
perturbations. This maximum disk mass is larger for magnetized disks
compared to those with $B$ = 0 (see Section \ref{sec:maxmass} and
equation [\ref{magmaxmass}]). These ideas can be tested through 
observations, as illustrated by the case of the disk surrounding 
the high-mass protostar Ceph A HW2 (Section \ref{sec:numerical}).

The inclusion of magnetic fields leads to significant modifications to
the prospects for the formation of giant planets through gravitational
instability. In order to form giant planets, the disk must be unstable
so that $Q_M$ is small, and the cooling time must be short enough.
Even in field-free disks, these coupled constraints limit the
formation of planets to take place at large radii.  The constraints
become tighter in the presence of magnetic fields (see equations
[\ref{sigmacon}] and [\ref{sigmaconnum}]).  Even in the event that
gravitational instabilities do occur, and cooling time scales are
short enough, the gravitationally bound fragments must loose large
amounts of magnetic flux in order to contract to planetary size
scales. This requirement leads to another constraint on the planet
formation process, namely, a lower bound on the electrical resistivity
(see equation [\ref{etacon}]). 

The results of this paper indicate that giant planet formation via 
gravitational instability is difficult and hence should occur rarely. 
Nonetheless, rare is not the same as never: Circumstellar disks that 
are sufficiently large (in radius) and massive could meet the 
constraints on this paper and support secondary formation.  Some of 
the planetary candidates that have discovered through direct imaging 
(e.g., Marois et al. 2008, Kalas et al. 2008, Lagrange et al. 2008) 
could represent examples where this process has taken place.

This paper represents an important step toward understanding the
effects of magnetic support in circumstellar disks. However, the
generalized stability parameter $Q_M$ derived here applies only to the
onset of gravitational instabilities, specifically, linear
perturbations with azimuthal wavenumber $m$ = 0. Spiral modes (with
$m\ne0$) can grow when axisymmetric perturbations are stable, so that
this work should be generalized to include higher wavenumbers.  In
addition, the long term fate of gravitational perturbations depends on
the nonlinear evolution of these magnetized disks, and hence fully 
time-dependent MHD calculations should be carried out (e.g., Inutsuka
et al. 2010). Since disks 
are expected to contain magnetic fields, these studies are vital to 
understanding both disk accretion and the possible formation of 
secondary bodies in star/disk systems.  

\acknowledgments

We would like to thank the Centro de Radioastronom\'\i a y Astrof\'\i sica,
UNAM (Morelia, Mexico), the Physics Department at U. C. San
Diego (La Jolla, CA), and Academia Sinica (Taiwan) for hosting and
helping to facilitate this collaboration. We would like to thank
Zhi-Yun Li and Frank Shu for useful discussions.  We also thank Mohsen 
Shadmehri for sharing unpublished calculations on the stability of 
cooling disks.
SL is supported by  Grant CONACyT 48901; MJC is supported by 
Grant NSC-95-2112-M-001-044; and FCA is supported by NASA Grant 
NNX07AP17G and NSF Grant DMS-0806756.

\appendix
\section{Appendix: Decay of magnetic fields in disks}

Equation (\ref{decay}) shows that the exponential decay rate of a
magnetic perturbation with radial wavenumber $k$ in an infinitesimally
thin disk is given by 
\be
\gamma \sim \eta\frac{k}{z_0} \, .
\ee
Thus, the characteristic diffusion length scale $\ell$ in an
infinitesimally thin disk is neither the radial wavelength of the
perturbation $\sim 1/k$ nor the disk thickness $\sim z_0$, but rather
the geometric mean of the two, $\ell\sim (z_0/k)^{1/2}$.  Since this
result is curious and at variance with the usual assumption that $\ell
\sim z_0$ in disks (e.g., see Parker~1979 and Zeldovich et al.~1983),
in the following we consider the expression of the decay rate derived
by Br\"auer \& R\"adler~(1988) and Krause~(1990) for a magnetic
perturbation in a disk of finite thickness (without the effects of
self-gravity, pressure, and rotation), and then we take the limit for
$z_0/\varpi \rightarrow 0$.

Consider a disk of finite thickness $2z_0$ and infinite radius,
and assume that $\eta$ is constant inside the disk. In this case,
the magnetic induction equation becomes 
\be
\pd{\bf B}{t}=-\eta\nabla\times (\nabla\times {\bf B}) 
\qquad \mbox{for $|z|<z_0$} \, .
\label{ind_int}
\ee
Next we assume a vacuum field outside the disk,
\be
\nabla \times {\bf B}=0 \qquad \mbox{for $|z|>z_0$},
\label{ind_ext}
\ee
and impose the condition of continuity of ${\bf B}$ at $z=\pm z_0$.

Any solenoidal field can be decomposed into a poloidal and toroidal
components defined by the scalar functions $P$ and $T$, respectively.
Note that for an axisymmetric system $P = \partial{A_\varphi}$/
$\partial{\varpi}$, $T=B_\varphi$, and the terms ``poloidal'' and
``toroidal'' assume their usual meaning. The magnetic field thus 
can be written in the form 
\be
{\bf B}=\nabla\times({\bf\hat e}_z\times \nabla P) + 
{\bf\hat e}_z \times \nabla T.
\ee
In terms of the defining scalars $P$ and $T$, equation~(\ref{ind_int}) 
separates into two pieces, 
\be
\pd{P}{t}=\eta\nabla P \qquad {\rm and} \qquad 
\pd{T}{t}=\eta\nabla T \, . 
\label{scal_int}
\ee
Equation~(\ref{ind_ext}) thus becomes
\be
\nabla^2 P=0, \qquad {\rm and} \qquad T=0 \, ,
\label{scal_ext}
\ee
The condition of continuity of ${\bf B}$ implies that $P$, $T$ and
$\partial P/\partial z$ must be continuous on $z=\pm z_0$.  The
general solutions of equation~(\ref{scal_int}) and (\ref{scal_ext})
that are regular at the origin can be expressed in terms of Bessel
functions of the first kind with the arguments $k\varpi$ and $k_z z$,
where $k$ and $k_z$ are the radial and vertical wavenumbers,
respectively.

\be
P(\varpi,\varphi,z,t)=\left\{
\begin{array}{ll}
\int_0^\infty [a_1(k)\sin(k_z z)+b_1(k)\cos(k_z z)]J_m(k\varpi)e^{im\varphi-\gamma t}\, dk & \mbox{if $|z|<z_0$} \\ \nonumber
\int_0^\infty c(k)e^{-k|z|} J_m(k\varpi)e^{im\varphi-\gamma t}\, dk & \mbox{if $|z|>z_0$}, \nonumber
\end{array}
\right.
\ee
and
\be
T(\varpi,\varphi,z,t)=\left\{
\begin{array}{ll}
\int_0^\infty [a_2(k)\sin(k_z z)+b_2(k)\cos(k_z z)]J_m(k\varpi)e^{im\varphi-\gamma t}\, dk & \mbox{if $|z|<z_0$} \\ \nonumber
0 & \mbox{if $|z|>z_0$}, \nonumber
\end{array}
\right.
\ee
where $k$ and $k_z$ are the radial and vertical wavenumbers, respectively,
$J_m(k\varpi)$ are Bessel functions of the first kind and
$a_1$, $a_2$, $b_1$, $b_2$ and $c$ are functions determined by the
initial field distribution and the conditions of continuity
on $z=\pm z_0$. The decay rate is
\be
\gamma=\eta(k_z^2+k^2),
\label{rate}
\ee
proportional to the inverse of the modulus of the vector wavenumber
${\bf k}=k\hat{\bf e}_\varpi+k_z\hat{\bf e}$.  The condition of
continuity of $\partial P/\partial z$ at $z=\pm z_0$ leads to the
relation
\be
(k_z^2-k^2)\tan (2k_z z_0)=2 k_z k.
\label{cont_cond}
\ee

The limit of an infinitesimally thin disk is recovered under the
following ordering: $z_0 \ll k_z^{-1} \ll k^{-1} \ll \varpi$ (the
disk thickness is much smaller than the vertical wavelength, that
in turn is much smaller than the radial wavelength). With these
approximations, equation~(\ref{cont_cond}) then gives $k\approx
k_z^2 z_0$ and equation~(\ref{rate}) becomes $\gamma\approx \eta
k_z^2\approx k/z_0$, as found in our case.

Note that in a thin disk, the size of axially symmetric magnetic
perturbations is limited vertically by the disk thickness, and,
horizontally by the radial extent of the perturbation.  The
perturbative calculation shows that the diffusion rate is not
determined by the smallest of these two scales ($z_0$, the disk
thickness) but by the geometric mean of the two. In a disk of finite
size, the relevant diffusion scale is determined, as expected, by
the modulus of the vector sum of the vertical and horizontal
wavenumbers; boundary conditions impose a relation between the two
wavenumbers that, in the thin disk limit, gives the derived result.

\end{document}